\documentclass[twocolumn,prb,a4,paper,superscriptaddress]{revtex4}

\usepackage[english]{babel}
\usepackage{psfrag}
\usepackage{hyperref}
\usepackage{graphicx}
\usepackage{amsmath}

\begin{document}

\title{Detection of Tiny Mechanical Motion by Means of the Ratchet Effect}

\author{Stefano Pugnetti}

\affiliation{NEST-CNR-INFM and Scuola Normale Superiore, Piazza dei Cavalieri 7, I-56126 Pisa, Italy}

\author{Yaroslav M.~Blanter}

\affiliation{Kavli Institute of Nanoscience, Delft University of Technology, Lorentzweg 1,  
             2628 CJ Delft, The Netherlands }

\author{Fabrizio Dolcini}

\affiliation{NEST-CNR-INFM and Scuola Normale Superiore, Piazza dei Cavalieri 7, I-56126 Pisa, Italy}

\author{Rosario Fazio}

\affiliation{NEST-CNR-INFM and Scuola Normale Superiore, Piazza dei Cavalieri 7, I-56126 Pisa, Italy}

\begin{abstract}
We propose a position detection scheme for a nanoelectromechanical resonator based on the ratchet effect. This scheme has an advantage of being a dc measurement. We consider a three-junction SQUID where a part of the superconducting loop can perform mechanical motion. The response of the ratchet to a dc current is sensitive to the position of the resonator and the effect can be further enhanced by biasing the SQUID with an ac current. We discuss the feasibility of the proposed scheme in existing experimental setups.
\end{abstract}

\maketitle

\section{Introduction}

In recent years the research on nanoelectromechanical systems (NEMS) developed as 
a broad and rapidly growing field\cite{Cleland,Clerk08}. The  low mass and small size of NEMS  make them 
excellent candidates for high-precision force\cite{Blencowe07}, mass\cite{Ekinci04}   and position 
detection\cite{Naik06,Li07,Flowers07,Blencowe07,Buks07,Regal08}, with a  sensitivity that is ultimately limited 
only by quantum mechanics. Measurements on SiN\cite{Lahaye04} and GaAs\cite{Knobel03} doubly-clamped beams coupled to single-electron transistors have   achieved the sensitivity of four times above the standard 
quantum limit. Moreover, NEMS have been recently envisaged as integrated elements into quantum information 
circuits, such as one\cite{Rabl04,Wang08,Hauss08} or several\cite{Zou04} qubits, opening up promising  
new prospectives also in the field of quantum information processing.

A crucial issue for exploiting the remarkable features of NEMS is the detection of their position. 
To this purpose two different techniques are currently applied: Optical detection  
based on coupling of an oscillator to a microwave cavity\cite{Arcizet06,Gigan06,Poggio07,Favero07,Pruessner08,Groeblacher08,Thompson08,Schliesser08} is very sensitive, although 
somewhat difficult to integrate in nanocircuits. On the other hand  traditional electrical detection\cite{Brown07}  
based on the oscillator back-action  offers the advantage that excitation and detection occur in the 
same circuit. However, this approach leads to lower sensitivity and does not allow an easy discrimination 
between electronic and mechanical excitations. 

Typically all existent detection schemes are based on high frequency measurements. In this article we 
suggest a new way to detect mechanical oscillations which avoids these difficulties and allows to detect 
the mechanical motion with a dc measurement. Our idea is based on the so-called ratchet effect~\cite{Reimann02}. 
In very simple terms the ratchet effect can be explained by considering a particle moving in an external 
periodic potential and subject to a periodic (or random) force with zero average. If the potential 
breaks spatial inversion symmetry, a directed motion is possible even if the system is unbiased. By detecting 
the directed motion of a particle it is in general possible to gain information on the oscillating force. Here we discuss in details the case when the 
nanomechanical resonator is a part of a SQUID loop. It is well known that SQUIDs are currently used as ultra 
sensitive   magnetic field detectors,  and that they represent  a solid-state implementation   of  qubits  
based on Josephson effect~\cite{Makhlin01}. A remarkable  attention in the last few years has also been devoted to 
coupling SQUIDs  to mechanical oscillators   for detection or cooling~\cite{Buks06, Xingxiang06, Xue07, 
Blencowe07, Wang08} and, very recently, the first detection of mechanical motion of a micromechanical resonator 
embedded in a dc SQUID was reported~\cite{etaki08}.

It has been shown~\cite{Zapata96} that SQUIDs can behave as ratchets, where the superconducting phase 
difference at the junction $\gamma$ and the Josephson potential play the role of the coordinate and the 
ratchet potential, respectively. When such systems are biased by an ac current, a constant drift in the 
phase drop occurs under appropriate conditions, causing a finite dc voltage.

In this work we propose to exploit the rectifying properties of these SQUID ratchets in order to detect the motion of a nanomechanical resonator; indeed a mechanical oscillator embedded in a SQUID circuit threaded by a magnetic flux produces an ac signal across the SQUID that plays the role of a zero-average driving force, affecting the dc characteristic of this ratchet circuit. We find that the current-voltage curve is qualitatively changed by the mechanical oscillations with respect to the case without a moving part; in principle certain features can be extracted that can only be attributed to the presence of mechanical oscillations, such as Shapiro steps and a non-monotonic dip-and-peak structure in the symmetric part of the curve coming from them. This reduces the problem of the detection of motion of a suitable class of NEMS to an accurate measurement of the dc current-voltage characteristic of this device, avoiding the difficulties in high-frequency ac measurements arising in other detection schemes.

The paper is organized as follows. In Sec.~\ref{Sec-2} we introduce the model and discuss the physically relevant scales. The results for the current-voltage characteristics are then presented; depending on physical parameters of the junctions employed, the adiabatic limit of a slowly oscillating resonator may be of relevance and it is presented in Sec.~\ref{Sec-3}. However, higher coupling efficiency may require slower time scales for the motion of the resonator, thus necessitating a more general discussion for arbitrary resonator frequency, which is done in Sec.~\ref{Sec-4}. We discuss the role of thermal fluctuations in Sec.~\ref{Sec-5}, and we then proceed with a critical assessment of the experimental realization of this scheme in existing setups (Sec.~\ref{Sec-6}) by comparing the numerical values of the relevant physical parameters. Concluding remarks and a thorough description of this detection technique are presented in Sec.~\ref{Sec-7}.

\begin{figure}[!tbh]
\centering
\includegraphics[width=\columnwidth]{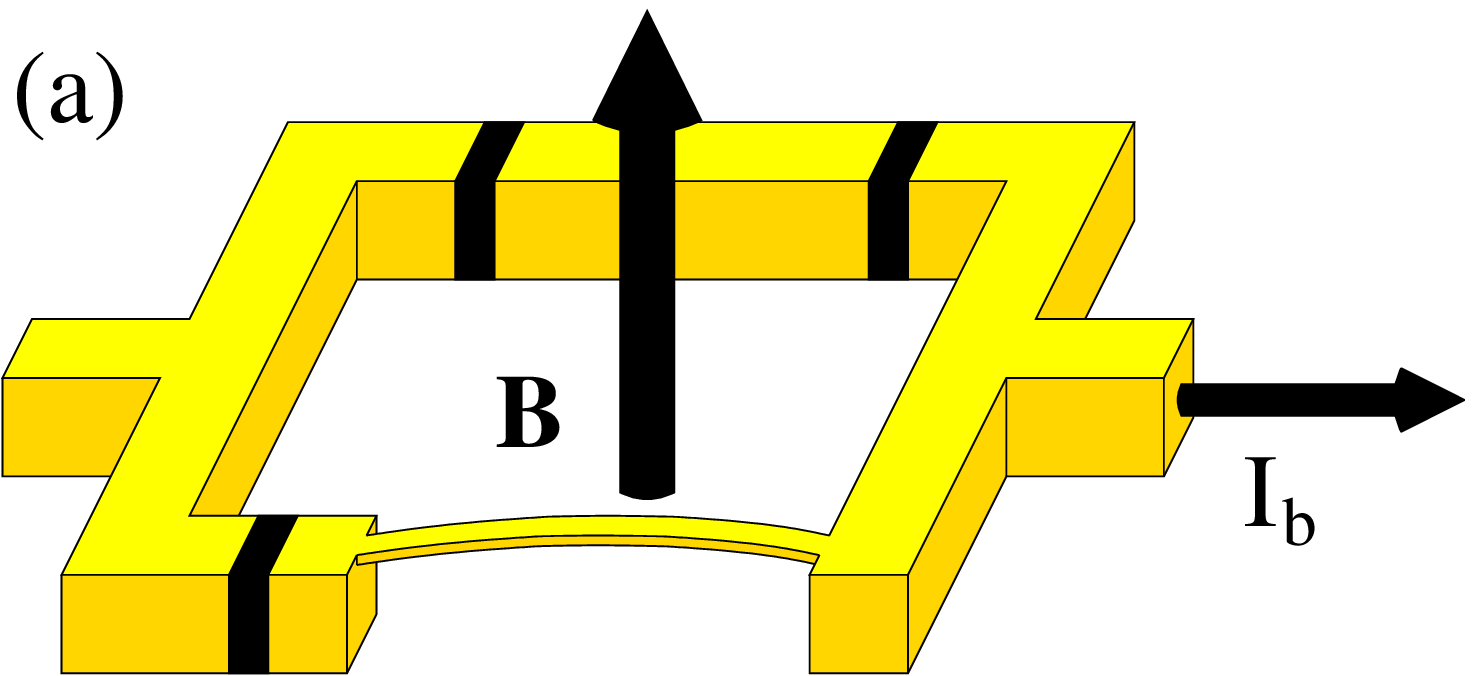}\\
\psfrag{1}{1}
\psfrag{B}{2}
\psfrag{3}{3}
\includegraphics[width=0.65\columnwidth]{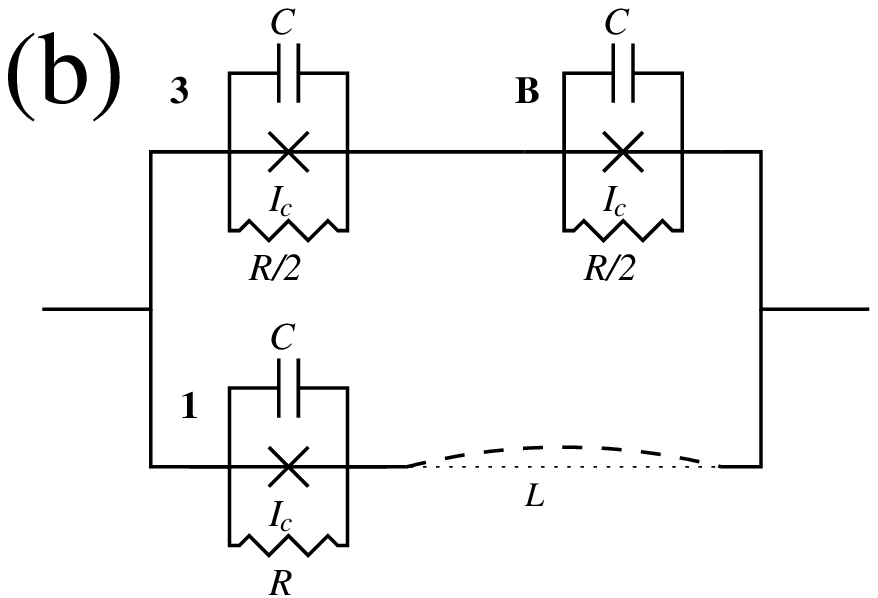}
\psfrag{gamma}{$\gamma$}
\psfrag{U}{$U$}
\includegraphics[width=0.3\columnwidth]{figure1c}
\caption{\label{fig:device}(a)-(b) A model and the corresponding scheme of the ratchet circuit considered here. Three Josephson junctions are coupled to a mechanical resonator oscillating in the plane of the circuit, threaded by a magnetic flux, and biased by a current. The junctions have equal critical currents and capacitances but different shunting resistances. (c) The potential $U(\gamma)$ in the absence of a moving part (solid line); the average slope in the decreasing part is larger in modulus than the average slope of the increasing part, as can be seen by comparing with a cosine curve (dashed line).}
\end{figure}

\section {The Model}
\label{Sec-2}
The circuit that we analyze is schematically depicted in Fig.~\ref{fig:device}.  
It consists of a Josephson junction (JJ)   and a mechanical resonator connected in parallel with two other JJs. 
The Josephson junctions are in the classical regime  and  they can be described with the RCSJ model, i.e.
\begin{equation}
  \frac{I_i}{I_c}=\sin\gamma_i+\frac{\hbar}{2e R_i I_c}\dot\gamma_i+\frac{\hbar C}{2eI_c}\ddot\gamma_i 
  \hspace{0.8cm} i=1,2,3 \;.
\label{i_i}
\end{equation} 
In Eq.(\ref{i_i}) the current flowing through each junction is denoted by $I_i$, 
and $\gamma_i$ is the related gauge-invariant phase difference. In Eq.~(\ref{i_i}) $C$ denotes the 
capacitance, and $R_i$ the resistances, chosen as $R_1=R$  and $R_2=R_3=R/2$ (see Fig.~\ref{fig:device}).  
For simplicity we assumed that the three junctions are characterized by the same critical current~$I_c$.
By considering  overdamped junctions, $\omega_{pl}R C\ll1$ ($\omega_{pl}=\sqrt{2eI_c/\hbar C}$ is their 
plasma frequency), the terms involving second-order derivatives of the phases can be neglected in Eq.~(\ref{i_i}). One can 
then show that, if $\gamma_2=\gamma_3$ holds at a certain time, such condition is maintained. We shall
henceforth set $\gamma_2=\gamma_3=\gamma/2$, whereas the phase 
$\gamma_1$ is related to the 
other ones by the equation
\begin{equation}
  \gamma_1-\gamma=2\pi\left(\frac\Phi{\Phi_0}+n\right)\qquad n=0,1,\ldots
\end{equation}
ensuring a vanishing total phase difference along the loop. 
In the previous equation $\Phi$ denotes the total magnetic flux threading the circuit and $\Phi_0$ is 
the elementary quantum of flux. Neglecting the self-inductance of the loop, the flux is determined 
by the total area of the circuit, which in turn depends on the position of the resonator: 
$\Phi=B(A+lX(t))$ ($l$ is the effective length of the resonator, $X(t)$ is the position of its 
center of mass and $A$ is the area of circuit when $X(t)=0$). In the following we will assume 
that the motion of the resonator is fixed externally and equals $X(t)=X_0\cos(\omega t)$ 
(see Sec.~\ref{Sec-6}). 
Indicating by $I_b$   the current biasing the device, current 
conservation $I_b(t)=I_1(t)+I_2(t)$ leads to the following equation of motion for the phase $\gamma$,
\begin{equation}
  \partial_\tau \gamma=-\frac{\partial U}{\partial \gamma}+\frac{I_b(\tau)}{I_c}+
                       \frac{\pi L B X_0}{\Phi_0} \Omega \sin(\Omega\tau) \;,
\label{eq:equation}
\end{equation}
where $\tau=\omega^* t$ is a dimensionless time variable, $\omega^*=eRI_c/\hbar$ and $\Omega=\omega/\omega^*$. 
We will consider external bias of the form $I_b(\tau)=I_{dc}+I_{ac}\cos(\Omega_{ac}\tau)$, with $I_{dc}$ denoting 
the dc component, and $I_{ac}$ and $\Omega_{ac}$ the amplitude and frequency of a monochromatic ac component.
The periodic function 
\begin{equation}
U(\gamma,f(\tau))=-2\cos(\gamma/2)-\cos(\gamma+\phi+f(\tau)) 
\label{U-def}
\end{equation}
is the potential leading to the ratchet effect as discussed by Zapata {\em et al}~\cite{Zapata96}.
In addition to the dimensionless external magnetic flux $\phi=2\pi AB/\Phi_0$, the potential~$U$ now includes also 
a time-dependent fluctuation part~$f(\tau)=(2\pi L B X_0/\Phi_0)\cos(\Omega\tau)$ which arises 
from the oscillations of the mechanical resonator.   
The coupling between the SQUID and the mechanical resonator can be parametrized by the dimensionless 
parameter
\begin{equation}
a=\frac{2\pi L B X_0}{\Phi_0}
\end{equation}
At finite temperatures the r.h.s.~of Eq.(\ref{eq:equation}) will also include a noise term. 

Depending on the value of the external flux, the potential term defined in Eq.(\ref{U-def}) breaks 
$\gamma$ inversion symmetry and therefore leads to the ratchet effect. In Ref.~\onlinecite{Zapata96}
this effect has been shown by biasing the SQUID with an oscillating current leading to a finite 
dc voltage. Here we show that the ratchet effect allows one to detect the oscillatory motion of the oscillator 
which enters the dynamics of the systems both in the potential through the fluctuating part of 
the flux and in the third term on the r.h.s of Eq.(\ref{eq:equation}).
We emphasize that, in order for the ratchet effect to appear, the system needs to be out of 
thermal equilibrium\cite{Reimann02}.  

The dc voltage drop across the circuit can be expressed  through  the Josephson relation  as
\begin{equation}
  V_0=\frac{\hbar}{2e}\left\langle\frac{d\gamma}{dt}\right\rangle=\frac{RI_c}{2}\langle 
      \partial_\tau\gamma \rangle \quad,
\end{equation}
where the brackets $\langle \ldots \rangle$ indicate time average (and possibly thermal average).
A non-vanishing voltage drop can be obtained when the ratchet potential is asymmetric in~$\gamma$, 
which occurs for values of the magnetic flux $\phi \neq  n\pi$. In particular one finds that, for 
sufficiently large values of $I_{ac}$, a finite dc voltage $V_0\neq0$ arises even in the presence of 
a vanishing average drive ($I_{dc}=0$). Furthermore, for $I_{dc} \neq 0$  the dc $V-I$ curve is not and 
odd function of the current. It is therefore convenient to analyze its symmetric (even) part 
defined as

\begin{equation}
  v_0^e \doteq \frac{V_0(I_{dc})+V_0(-I_{dc})}{2RI_c} \;.
\label{even-comp}
\end{equation}

Mechanical oscillations lead to two effects in the dynamics of the SQUID. 
On one hand a term qualitatively similar to 
an  ac bias arises, as shown by the last term in the r.h.s of Eq.(\ref{eq:equation}). Secondly,  
a fluctuating term $f(\tau) \neq 0$ appears in the potential, affecting the symmetry of the 
ratchet potential as a function of $\gamma$. In the presence of the resonator   the analysis 
is thus more subtle. As we shall see in Sec.~\ref{Sec-3}, the  fluctuations of the potential tend  to wash out the 
ratchet effect; although for some specific values of the parameters one can still obtain 
$V_0(I_{dc}=0)\neq 0$. 
In most cases the ratchet effect survives only in some non-trivial features in the symmetric part 
of the $V-I$ curve, which we will mainly focus on in the rest of the paper.

\begin{figure}[!t]
\centering
\psfrag{v0}{\Large$v_0$}
\psfrag{i0}{\Large$I_{dc}/I_c$}
\psfrag{v0e}{\Large$v_0^e$}
\psfrag{f=p/12}{$\phi=\pi/12$}
\psfrag{f=p/2}{$\phi=\pi/2$}
\psfrag{f=2p/3}{$\phi=2\pi/3$}
\psfrag{f=3p/2}{$\phi=3\pi/2$}
\includegraphics[width=\columnwidth]{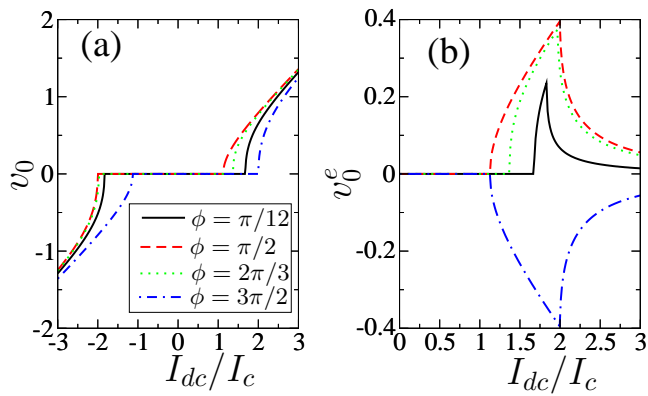}
\caption{\label{fig:noa}(a) Dimensionless dc voltage drop $v_0=V_0/RI_c$ as a function of the dimensionless bias current $I_{dc}/I_c$, for different values of the flux. No oscillating mechanical part is present ($a=0$). (b) The even part $v_0^e$ of the same curves plotted 
for positive values of $I_{dc}$ only. $v_0^e$ is zero for $\phi=n\pi$ and $v_0^e(-\phi)=-v_0^e(\phi)$.}
\end{figure}

 Apart from some specific cases the equation of motion of the SQUID, (\ref{eq:equation}),
does not allow for an analytic solution. Most of our results are based on a numerical 
integration of the equation of motion. As we are interested in using the SQUID as a 
detector of the mechanical oscillations, it is appropriate to first briefly discuss the case 
where no oscillator is present, in order to highlight the difference due to the mechanical 
motion. When the SQUID is biased by a dc current [$I_b(\tau) = I_{dc}$] and there is no 
resonator ($a=0$) the voltage drop can straightforwardly be computed as $V_0=4\pi (R I_c/2) /T$, 
where $4\pi$ is the   period of the potential (\ref{U-def}), and 
\begin{equation}
\label{eq:exact}
       T=\int_0^{4\pi}\frac{d\gamma}{\displaystyle I_{dc}/I_c-\sin(\gamma/2)-\sin(\gamma+\phi)}\qquad 
\end{equation}
is the time required for the phase to span such an angle.
The resulting $V-I$ curve of the device is shown in Fig.~\ref{fig:noa}(a). For an 
asymmetric potential ($\phi\neq n\pi$) the curves are not odd in the current, and a 
non vanishing even component (\ref{even-comp}) arises [see Fig.~\ref{fig:noa}(b)]. 
This component will be shown to be affected by the presence of the mechanical resonator. 
As the flux varies, the curve~$v_0(I_{dc})$  shifts to the left and then to the right with 
a period of $2\pi$; this reflects in the amplitude and sign of $v_0^e$. Indeed, from 
Eq.~(\ref{eq:exact}) one can show that $v_0^e(-\phi)=-v_0^e(\phi)$ and $v_0^e(\phi=0)=0$.

\section{Slowly oscillating resonator - the adiabatic limit}
\label{Sec-3}
We now discuss the role of the resonator.
Before addressing the full numerical solution of Eq.~(\ref{eq:equation}), we discuss here the limiting 
situation in which $\Omega \ll 1$, i.e. the oscillator's frequency is much smaller than the plasma frequency of the 
SQUID. As we shall see, this regime is indeed suggested by the typical experimental conditions, where   
$\Omega \sim 10^{-3}$, indicating that the mechanical oscillations are much slower than the dynamics of the 
junctions. Moreover the same condition can be realized for the frequency of the ac component of the bias 
current. Under these conditions, the equation of motion can be solved within the adiabatic approximation. 
One can determine   $v_0(I_{dc})$ by evaluating, for a fixed position of the oscillator~$X(\tau)$ and value of 
the bias current $I_b(\tau)$, the time~$T(\tau)$ required for the phase to span an angle  $4\pi$. Averaging 
over $\tau$ one then obtains
\begin{eqnarray}
      v_0 &=& \frac12\langle 4\pi/T(\tau) \rangle= \label{eq:adiabatic}\\
      &=& \int_0^{2\pi} \left[ \int_0^{4\pi}\frac{d\gamma}{ I_b(\theta)/I_c- \partial_{\gamma}U(\gamma,f(\theta))
      +\frac{a\Omega}{2}\sin(\theta)} \right]^{-1} d\theta \nonumber
\end{eqnarray}
Equation (\ref{eq:adiabatic}) will be analyzed in two different cases in which the bias current is constant 
or has also an additional monochromatic ac component.

\subsection{dc bias current}
\begin{figure}[!t]
\centering
\psfrag{v0}{\Large$v_0$}
\psfrag{i0}{\Large$I_{dc}/I_c$}
\psfrag{v0e}{\Large$v_0^e$}
\psfrag{a=0}{$a=0$}
\psfrag{a=0.01}{$a=0.01$}
\psfrag{a=0.1}{$a=0.1$}
\psfrag{a=1}{$a=1$}
\psfrag{a=20}{$a=20$}
\includegraphics[width=\columnwidth]{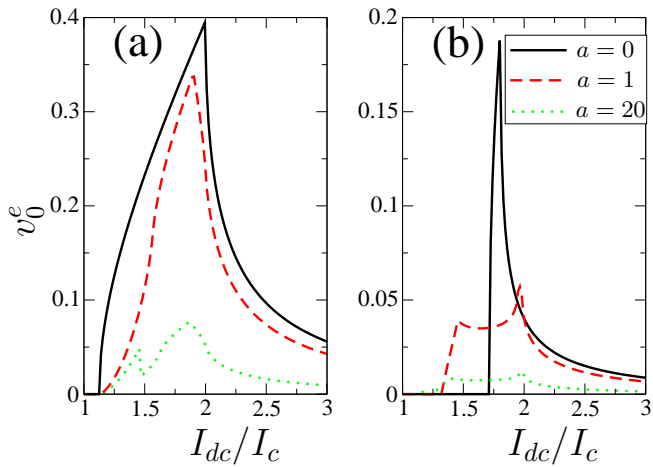}
\caption{\label{fig:noi1}$v_0^e(I_{dc})$ curves for different values of the coupling parameter $a$. The two plots refer to different values of the flux: (a) $\phi=\pi/2$  and (b) $\phi=\pi/20$. While for small coupling one recovers the curves of Fig.~(\ref{fig:noa}), for high coupling the features are washed out. The effect is more pronounced for small flux bias.}
\end{figure}

We start by analyzing the situation of a dc bias current, $I_b(\tau)=I_{dc}$. In Fig.~\ref{fig:noi1}(a) 
and \ref{fig:noi1}(b) the even component (\ref{even-comp}) of the voltage is plotted as a function 
of $I_{dc}$, for different values of the coupling strength $a$ and  flux $\phi$. For small coupling 
$a \ll 1 $, we recover the  exact results discussed in Sec.\ref{Sec-2} [see Fig.~\ref{fig:noa}(b)]. On the other side,
at high values of the coupling the main effect of the mechanical oscillations is to wash out the effects 
of the asymmetry of the ratchet potential. This is due to the fact that in this system the resonator 
oscillations also affect the ratchet potential via the term~$f(t)$.
Notice that the disappearance of asymmetry occurs at lower values of $a$ for a smaller flux $\phi$, as 
one can see by comparing the dashed-dotted curves in Fig.~\ref{fig:noi1}(a) and \ref{fig:noi1}(b). Thus the effects
of the coupling to the mechanical resonator are more dramatic for $\phi\sim n\pi$. Note, however, that for the special values $\phi=n \pi$, where $n$ is 
an integer, one still finds $v_0^e=0$, as can be shown by a change of integration variables  
$\gamma\rightarrow-\gamma$ and $\theta\rightarrow\pi+\theta$ in Eq.~(\ref{eq:adiabatic}) with 
$I_b(\tau)\equiv I_{dc}$.  We notice also that the experimentally accessible values of $a$ are very small. 
Although in Sec.\ref{Sec-6} possible operative ways to enhance the coupling $a$ will be proposed, these 
results lead to conclude that, in the presence of a purely dc bias $I_{dc}$, it is difficult to gain information 
on a realistic   mechanical resonator. A possible way out is to inject current with a monochromatic ac 
component with frequency $\omega_{ac}/2\pi$, small enough for the adiabatic approximation still to hold. 
This is discussed in the next subsection.

\subsection{ac monochromatic bias}

In this case  some new effects that can  be clearly attributed to the  mechanical resonator   
indeed arise. One might naively expect that  such effects simply originate from resonances at 
specific values of $\omega_{ac}$  related to $\omega$, yielding an enhancement of the effects due 
to the mechanical oscillations. However, as we shall discuss below, the situation is more 
complex and the overall features of the current-voltage characteristics originate from different 
mechanisms.

\begin{figure}[!t]
\centering
\psfrag{v0}{\Large$v_0$}
\psfrag{i0}{\Large$I_{dc}/I_c$}
\psfrag{v0e}{\Large$v_0^e$}
\includegraphics[width=\columnwidth]{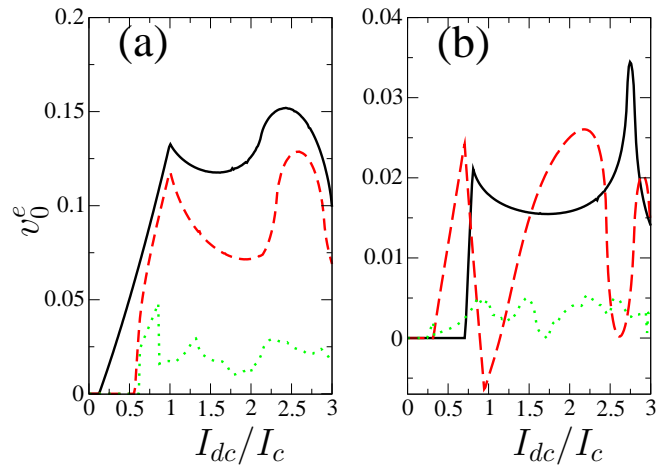}
\caption{\label{fig:i1}$v_0^e(I_{dc})$ curves for $I_{ac}=I_c$, $\Omega_{ac}=2\Omega/3$ and different values of the coupling parameter $a$ (the same as in Fig.~\ref{fig:noi1}). The two plots refer to different values of the flux bias: (a) $\phi=\pi/2$  and (b) $\phi=\pi/20$. Again for small coupling the curves are indistinguishable from the uncoupled ones ($a=0$); however for high couplings now a richer structure appear and its set-on occurs at a smaller coupling for $\phi\sim n\pi$.}
\end{figure}

We assume $I_b(\tau)=I_{dc}+I_{ac}\cos(\Omega_{ac}\tau)$ ($\Omega_{ac}=\omega_{ac}/\omega^*$) and we restrict to 
the case $ \Omega_{ac} =\Omega q/p$, with $q$ and $p$ integers with no common divisor apart from 1. 
A straightforward calculation allows to rewrite  Eq.~(\ref{eq:adiabatic}) as 
\begin{eqnarray}
v_0&=&\int_0^{2\pi}d\theta\left\{\int_0^{4\pi}d\gamma\left[
\frac{I_{dc}}{I_c}+\frac{I_{ac}}{I_c}\cos(q\theta)-\sin\left(\frac\gamma2\right)-
\right.\right. \nonumber \\
& & -\left.\left.
\sin[\gamma+\phi+a\cos(p\theta)]+\frac{a\Omega}{2}\sin(p\theta)
\right]^{-1}\right\}^{-1}
\end{eqnarray}
and to determine how the ac bias affects the $V-I$ curves of Fig.~\ref{fig:noi1}. The result is 
shown in  Fig.~\ref{fig:i1} for two different values of the flux.

\begin{figure}[!t]
\centering
\psfrag{i0}{\Large$I_{dc}/I_c$}
\psfrag{v0e}{\Large$v_0^e$}
\psfrag{O1=2/3O}{$\Omega_{ac}=2/3\Omega$}
\psfrag{O1=4/5O}{$\Omega_{ac}=4/5\Omega$}
\psfrag{O1=2/5O}{$\Omega_{ac}=2/5\Omega$}
\includegraphics[width=\columnwidth]{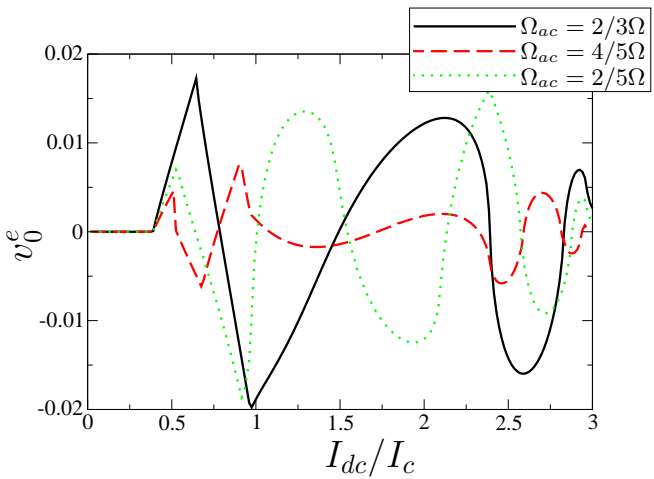}
\caption{\label{fig:several_omega1}$v_0^e(I_{dc})$ for zero bias flux and different values of the ratio $\Omega_{ac}/\Omega$. Here $a=1$, while for $a=0$ the curves should be flat ($v_0^e\equiv0$).}
\end{figure}

The even component $v_0^e$ is suppressed for strong coupling, similarly to the dc bias case. However, a richer structure  due to the interplay 
of the two components of the driving bias is observable. Another noteworthy difference with respect to the dc case is that, at small values of the dimensionless flux, the suppression occurs at higher values of the parameter $a$, as one can observe by comparing the curves of Fig.~\ref{fig:i1}(b) with Fig.~\ref{fig:noi1}(b). This indicates that the presence of an ac current makes the 
even component of the characteristic curve much more robust to the coupling with the resonator. Such difference is particularly striking at $\phi=n\pi$ ($n=0,\pm1,\ldots$). For 
these particular values of flux $v_0^e$ vanishes  for a purely dc bias current, whereas a finite value of $v_0^e$ is predicted for an ac current. Indeed the argument used to prove that $v_0^e\equiv0$ if $\phi=n\pi$ (a suitable change in the integration variables) does not apply when $I_{ac} \neq 0$, unless the frequency assumes specific values, namely those for which $p$ and $q$ are both odd integers. One can see the effects of the two combined oscillations in Fig.~\ref{fig:several_omega1}. Notice that these curves cannot be due to an ac bias \emph{alone}, for if $a=0$ and $I_{ac}\neq0$ the same argument of the previous section would apply, and the $V-I$ curve at $\phi=n\pi$ would be odd.


These results suggest a possible operative detection protocol for the tiny motion of a mechanical 
resonator integrated in a SQUID ratchet circuit: one may first measure the   $V-I$ curve of the circuit 
biased by a purely dc current, for different values of magnetic flux $\phi$. This would allow to determine 
values of $\phi$ such that the $V-I$ curve is completely odd. Then, by adding an ac term to the bias 
current, without changing the magnetic field, a non-vanishing even component would arise in the $V-I$ curve 
with varying the frequency of the ac bias.

\section{Arbitrary resonator frequency}
\label{Sec-4}

In many situations the value of the coupling parameter~$a$ is rather small (see Sec.~\ref{Sec-6}); for improving the coupling efficiency one then could in principle study devices with a higher value of the frequency $\Omega$, so that the coupling term $a\Omega\cos(\Omega\tau)/2$ can be increased. Under these circumstances the adiabatic approximation may not hold anymore and one should compute the current-voltage characteristic starting from the complete numerical solution of Eq.~(\ref{eq:equation}).

\begin{figure}
\centering
\psfrag{i0}{\Large$I_{dc}/I_c$}
\psfrag{v0}{\Large$v_0$}
\psfrag{O=1}{$\Omega=1$}
\psfrag{O=0.5}{$\Omega=0.5$}
\psfrag{O=0.1}{$\Omega=0.1$}
\psfrag{O=0.01}{$\Omega=0.01$}
\psfrag{a=1}{$a=1$}
\psfrag{a=0.5}{$a=0.5$}
\psfrag{a=0.1}{$a=0.1$}
\psfrag{a=0.01}{$a=0.01$}
\includegraphics[width=\columnwidth]{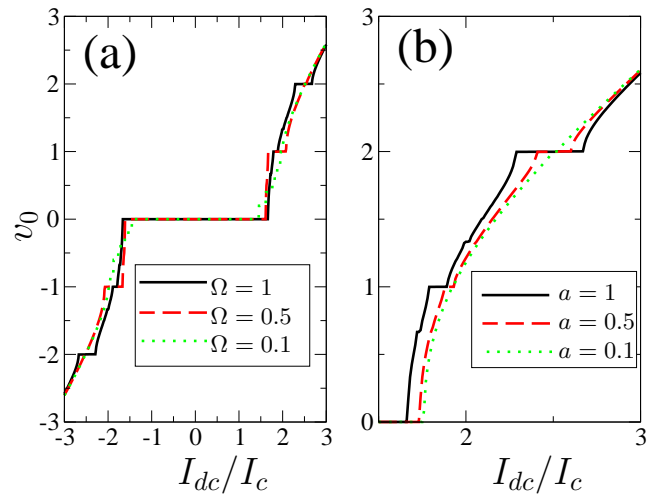}
\caption{\label{fig:shapiro}(a) Exact $v_0$-$I_{dc}$ curves for different frequencies $\Omega$ of the mechanical resonator. Here $a=1$, $I_{ac}=0$ and $\phi=0$. Even though no ac bias current is supplied, the system exhibits Shapiro steps, whose heights is related to $\Omega$. (b) The same but for a fixed frequency $\Omega=1$ and different values of the coupling $a$. The width of Shapiro steps approaches zero for small values of $a$.}
\end{figure}
The results are shown in Fig.~\ref{fig:shapiro}(a), where the $V-I$ curves at strong coupling are plotted, in the absence of the ac bias current, at vanishing flux and for different values of the dimensionless frequency $\Omega$. One can clearly recognize the typical shape of the $V-I$ curve for ac biased Josephson junctions, characterized by Shapiro steps. However, while these features are usually attributed to an ac bias current, here they originate from the mechanical oscillations. Notice that the height of the steps varies with the proper frequency of the resonator, as it happens for Shapiro  steps in isolated Josephson junctions. This suggests that in principle the frequency of the resonator can be directly read out by inspection of the height of the steps in the $V-I$ curve of the system. The width of the steps, instead, is usually related to the amplitude $a$ of the ac biasing signal. This is illustrated in Fig.~\ref{fig:shapiro}(b), where one can see that, as $a$ approaches zero, the steps become narrower and the $V-I$ curve becomes smoother.
\begin{figure}[!t]
\centering
\psfrag{i0}{\Large$I_{dc}/I_c$}
\psfrag{v0e}{\Large$v_0^e$}
\psfrag{a=1}{$a=1$}
\psfrag{a=0.5}{$a=0.5$}
\psfrag{a=0.1}{$a=0.1$}
\psfrag{a=0.01}{$a=0.01$}
\includegraphics[width=\columnwidth]{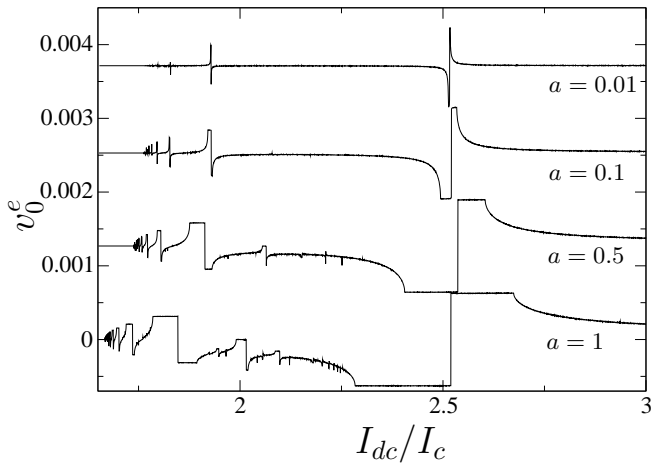}
\caption{\label{fig:even}The even component $v_0^e$ of the voltage drop as a function of the dc bias current $I_{dc}$  at $\phi=0$ and $\Omega=1$ for different values of the coupling $a$ (the curves are displaced vertically). The amplitude of the features is the same for all couplings, but their widths becomes smaller for smaller couplings.}
\end{figure}

Important differences emerge with respect to the adiabatic limit. In particular the even component of the $V-I$ curve does not vanish for $\phi=n\pi$ and $I_{ac}=0$, and exhibits an interesting structure: ranges where the voltage is independent of $I_{dc}$ are separated by sequences of maxima and minima, as shown in Fig.~\ref{fig:even}. Interestingly, the amplitude characterizing these features in the even component of the $V-I$ curve seems roughly independent of the coupling parameter $a$. This represents a promising effect for the detection of realistic mechanical oscillators. However, as it is already the case for the whole curve~$v_0(I_{dc})$ (Fig.~\ref{fig:shapiro}), the width of the different peaks is suppressed for small coupling.
\begin{figure}[!b]
\centering
\psfrag{i0}{\Large$I_{dc}/I_c$}
\psfrag{v0e}{\Large$v_0^e$}
\psfrag{a=1}{$a=1$}
\psfrag{a=0.5}{$a=0.5$}
\psfrag{a=0.1}{$a=0.1$}
\psfrag{a=0.01}{$a=0.01$}
\includegraphics[width=\columnwidth]{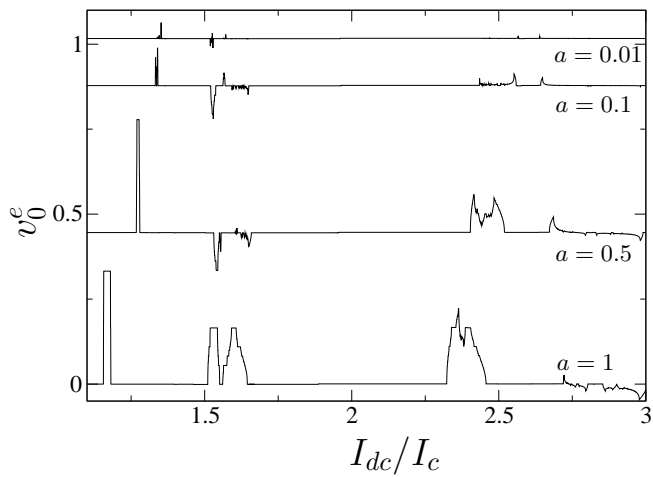}
\caption{\label{fig:even_i1} The even component $v_0^e$ of the voltage drop as a function of the dc bias current $I_{dc}$ at $\phi=0$ and $I_{ac}=I_c$, for different values of the coupling $a$. Here $\Omega=1$ and $\Omega_{ac}=0.67$.}
\end{figure}

The effect of an ac bias is to enhance the amplitude of these features; indeed the ratio between the heights of the features 
in Fig.~\ref{fig:even_i1} and those in Fig.~\ref{fig:even} is about~$50$. The numerical solution indicates that, 
in the absence of a mechanical oscillator ($a=0$), the $V-I$ curve is odd ($v_0^e\equiv0$), as in the adiabatic 
limit. We conclude that at $\phi=n\pi$ a detected signal has to be attributed to  the mechanical resonator only.  
Even though the main difficulty at low coupling still remains,  these effects may be enhanced by applying an ac bias.

\section{Finite Temperatures}
\label{Sec-5}
\begin{figure}[!t]
  \psfrag{v0e}{\Large$v_0^e$}
  \psfrag{i0}{\Large$I_{dc}/I_c$}
  \includegraphics[width=\columnwidth]{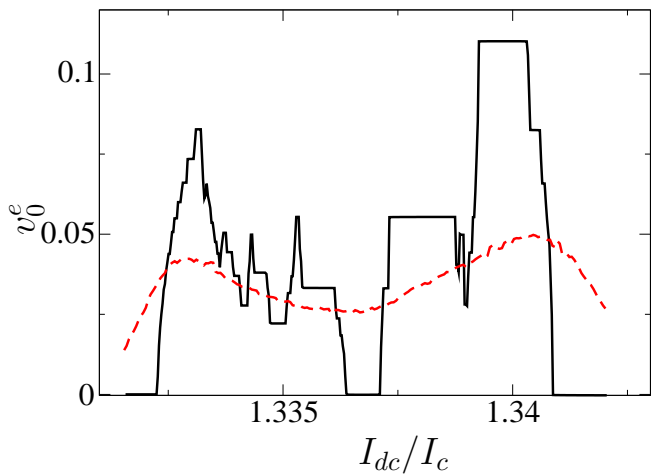}
  \caption{\label{fig:T}Some details of Fig.~\ref{fig:even_i1} for $a=0.1$, $I_{ac}=1$, $\Omega=1$ and $\Omega_{ac}=0.67$; the curves correspond to different temperatures: $\sqrt{\Theta}=0$ (solid line) and $\sqrt{\Theta}=10^{-4}$ (dashed line).}
\end{figure}

In this section we discuss the effects of thermal fluctuations on the dc voltage drop of the system. To be consistent with the model  adopted so far, we shall focus on the classical regime  and account for finite temperature by adding a stochastic term (white noise) in Eq.~(\ref{eq:equation}). These fluctuations are associated to two sources of dissipation: the finite resistance $R/2$ of the circuit, giving rise to electrical fluctuations in the bias current, and the finite quality factor $Q$ of the mechanical resonator, related to a random force producing mechanical fluctuations. Electrical fluctuations are taken into account by adding a term 
\begin{equation}
\Delta I_b(\tau)= I_c \sqrt{\frac{4ek_BT}{\hbar I_c}}\xi(\tau) \quad,
\end{equation}
to the bias current $I_b(\tau)$, where $\xi(\tau)$ is a white-noise process with zero average and  correlation function $\langle\xi(\tau)\xi(\tau')\rangle=\delta(\tau-\tau')$. Mechanical fluctuations may be introduced via a stochastic force
\begin{equation}
 F(t)=\sqrt{\frac{2m\omega k_BT}{Q}}\eta(t)
\end{equation}
acting on the resonator. Here $m$ is the mass of the mechanical oscillator and $\eta(t)$ another white-noise process, statistically independent of $\xi$. In the previous sections we assumed the motion of the mechanical oscillator to be undamped ($Q=\infty$), so that the mechanical noise should be neglected. However for realistic devices this treatment is still a good approximation if 
\begin{equation}
\label{eq:noiseless}
k_BT\ll Q\frac 12 m\omega^2x_0^2
\end{equation}
with $x_0$ denoting the amplitude of the oscillation. For high $Q$ there is a regime in which temperature is low enough to fulfill condition (\ref{eq:noiseless}), but high enough to fulfill $k_BT\gg\hbar\omega$, so that the behavior is neither quantum nor noisy. In the following we shall assume that the mechanical noise can be neglected, and analyze Eq.~(\ref{eq:equation})  with 
\begin{equation}
y=a\Omega/2\sin(\Omega\tau)+\frac{I_{dc}}{I_c}+\frac{I_{ac}}{I_c}\cos(\Omega_{ac}\tau)+\sqrt{\Theta}\xi(\tau)
\end{equation}
where $\Theta=T/T_0$ and $T_0=\hbar I_c/4ek_B\simeq 12\mathrm{K}\cdot I_c/1\mu\mathrm{A}$. As one may expect, for a certain value of the coupling parameter $a$ there is a value of $\Theta$ above which the peculiar features of the characteristic curve are lost. For determining the temperature scale at which this happens, one has to know $T_0$, which is set by the value of the critical current. Devices with large critical currents are in principle the best for this detection scheme, since they will be less affected by temperature. For a critical current as large as 1mA, the temperature scale is $10^4$K. As an example, consider Fig.~\ref{fig:T}, which is a detail of the $v_0^e(I_{dc})$ curve in Fig.~\ref{fig:even} for $a=0.1$, compared with its finite temperature counterpart. Thermal fluctuations tend to wash out the peculiar features of the dc characteristic; the temperature scale at which this happens is $0.1$mK for a high-$I_c$ device, corresponding to $\Theta\approx10^{-8}$.


\section{Experimental perspectives}
\label{Sec-6}
By now top-down fabrication techniques can produce single-cristal doubly-clamped beams whose typical dimensions are $1\mu\mathrm{m}\times0.1\mu\mathrm{m}\times0.1\mu\mathrm{m}$ and whose frequency ranges from tens of MHz\cite{Cleland96} to 1 GHz\cite{Huang03}; using Silicon Nitride or Carbide, the density is about $3\times10^3\mathrm{kg/m^3}$ and hence typical masses are of the order of $10^{-17}\mathrm{kg}$. Lower values can be obtained using carbon nanotubes, which can also have higher oscillation frequencies due to their high bulk modulus\cite{Sapmaz03}. The oscillation can be induced by driving the resonator either mechanically (with a piezo element applied on the substrate and driven by an ac signal) or electrically (for instance by capacitively coupling the resonator with a gate electrode); a $1\mu$m-long beam can be easily excited up to oscillation amplitudes of 1nm. To estimate the corresponding coupling parameter $a$ one has to keep in mind that the magnetic field causing the coupling cannot be too large, otherwise the device loses superconductivity; a typical upper bound is $B<0.1$T. With this values, one gets $a\approx0.1$, the value used for the plots in Fig.~\ref{fig:T}.

The intensity of the back-action on the beam can be estimated by the Lorentz force acting on it in virtue of the current flowing through it: $F_l=LIB\simeq 1\mu\mathrm{m}\times1m A\times0.1T=10^{-10}N$; this is to be compared with  $m\omega^2x_0\simeq10^{-17}\mathrm{kg}\times4\pi^2(1\mathrm{GHz})^2\times10^{-9}\mathrm{m}=4 \times10^{-7}N$. (At such amplitudes the force may be no longer linear in the displacement and it would probably be larger than this estimate). The back action can thus be neglected as a first approximation.

As for the adiabatic approximation, the value of $\Omega=\hbar\omega/eRI_c$ depends on the resistance of the junction. For tunnel junctions the low-temperature resistance is due only to quasi-particle tunneling and increases exponentially with decreasing temperature; at temperatures close to the critical temperature the resistance can be estimated with the normal state resistance and $I_cR_n$ can be as large as 1mV, corresponding to $\omega^*\simeq10^{12}$Hz. In such situations the adiabatic approximation could apply. However the resistance of a Josephson junction can vary greatly and thus the validity of the adiabatic approximation depends on the device.

\section{Conclusions}
\label{Sec-7}

A SQUID ratchet circuit can in principle be used to detect the motion of a nanomechanical resonator integrated in it by means of dc measurements only. The characteristic current-voltage curve of this circuit exhibits Shapiro steps whose heights can be used to determine the frequency of the mechanical oscillations. Once this frequency is known, either directly with this dc technique or with others, one can bias the device with an ac signal whose frequency should be chosen in order to improve the effects of the mechanical oscillations; we suggest to use an ac frequency of 2/3 the estimated frequency of mechanical oscillations. Now, by varying the magnetic field through the device, the symmetric part of the characteristic curve should evolve periodically (the period being set by the flux quantum); the amplitude of the curve will be maximum for frustrated values of the flux ($\Phi_e=(n + 1/2)\Phi_0$) and minimum for unfrustrated values ($\Phi_e=n\Phi_0$). If there were no mechanical oscillations in the circuit, then the symmetric part of the characteristic curve should be zero for unfrustrated flux bias, since the potential experienced by the superconducting phase would be symmetric; everything that can be seen in this case is due to the interplay between the ac driving field and the signal due to the mechanical oscillations of the integrated nanoresonator. The shape of the characteristic curve can be compared with the finite temperature results we illustrated in this work (see for instance Fig.~\ref{fig:T}). We emphasize that no ac measurement is required for this detection scheme.

We acknowledge very fruitful discussions with Herre van der Zant, and Menno Poot. The work was supported 
by SNS05 of Scuola Normale Superiore and by Netherlands Foundation for Fundamental Research on Matter (FOM). S.~Pugnetti acknowledges the Kavli Institute of Nanoscience for the kind hospitality, and F.~Dolcini the Italian MIUR "Rientro dei Cervelli" Program for financial support.

\end{document}